\documentstyle[prb,twocolumn,aps,psfig,amssymb,floats]{revtex}

\newcommand{\partabl}[2]{\frac{{\rm \partial} #1}{{\rm \partial} #2}}

\begin{document}
\draft

\twocolumn[\hsize\textwidth\columnwidth\hsize\csname @twocolumnfalse\endcsname


\title{Quantum phonons and the charge density wave transition
temperature: a dynamical mean field study} 
\author{Stefan Blawid and Andrew J.~Millis}
\address{Center for Materials Theory\\ Department of Physics \&
Astronomy, Rutgers University\\ 136 Frelinghuysen Road, Piscataway, NJ
08854} 
\date{\today}

\maketitle


\begin{abstract}
We use the dynamical mean-field method to calculate the charge density
wave transition temperature of the half-filled Holstein model as
function of typical phonon frequency in the physically relevant
adiabatic limit of phonon frequency $\Omega$ much less than electron
bandwidth $t$. Our work is the first systematic expansion of the
charge density wave problem in $\Omega/t$. Quantum phonon effects are
found to suppress $T_{\rm co}$ severely, in agreement with previous
work on one dimensional models and numerical studies of the dynamical
mean field model in the extreme antiadiabatic limit ($\Omega \sim
t$). We suggest that this is why there are very few CDW systems with
mean-field transition temperatures much less than a typical phonon
frequency.
\end{abstract}



\vskip2pc]

\section{Introduction}
\label{sec:intro}

The present-day understanding of electron-lattice interactions in
metals is based on the Migdal-Eliashberg (ME) theory\cite{migdal58},
which exploits the smallness of typical phonon frequencies, $\Omega$,
and temperatures, $T$, relative to typical electronic energies,
$t$. This theory may be regarded as an expansion in powers of the
adiabatic parameter $\gamma = {\rm max}(\Omega,T)/t$, in which only
the leading nontrivial term is retained. In many cases the leading
term is of order $\gamma^0$ and thus within ME theory most quantities
of physical interest are independent of ion mass. A notable exception
is the superconducting transition temperature $T_c$, which is
controlled by a logarithmic divergence with an upper cutoff set by the
phonon frequency, and involves an interaction (the Coulomb
pseudopotential) which has a logarithmic suppression with a lower
cutoff set by the phonon frequency. The resulting, rather
complicated, dependence of $T_c$ on ion mass was determined by
McMillan\cite{mcmillan68}.

A series of recent experimental papers reporting isotope effects
(i.e.~ion-mass dependence) for a variety of electronic properties
\cite{lacamno,laprcamno} motivates re-examination of this issue. Of
particular interest here are reports of a pronounced increase of the
charge ordering transition temperatures of various members of the `CMR
manganite' family of manganese perovskites when $\rm ^{18}O$ is
substituted for $\rm ^{16}O$.\cite{isaac98,zhao98} This leads us to
re-examine the theory of charge density wave (CDW) instabilities in
order to determine how the transition temperature $T_{\rm co}$ is
affected by ion mass. A theory applicable to materials of current
interest must go beyond the Migdal-Eliashberg approximation, must
apply to the case of $d=3$ spatial dimensions and must be able to
treat the physically relevant limit $\gamma \ll 1$.

The $\Omega$-dependence of $T_{\rm co}$ has been studied. In one
spatial dimension pioneering numerical work of Hirsch and
Fradkin\cite{hirsch83} followed by analytical studies by Bourbonnais
and Caron\cite{bourbonnais89} found a strong suppression of the zero
temperature CDW order parameter with increasing $\Omega$. In $d=2$
\cite{marsiglio90,noack} and $d = \infty$\cite{freericks93} the
quantum phonon problem has been attacked by a direct numerical
approach using quantum Monte Carlo (QMC) methods. However, numerical
issues related to the mismatch between phonon and electron frequency
scales however restricted the QMC
studies\cite{marsiglio90,noack,freericks93} to the antiadiabatic
limit, $\Omega \sim t$. Also, in Refs.
\onlinecite{marsiglio90,noack,freericks93} the isotope effect on the
CDW transition temperature was not studied in detail, although values
for $T_{\rm co}$ for a given phonon frequency were calculated. In $d =
\infty$ and for weak coupling strengths the CDW transition temperature
is strongly reduced with decreasing ion mass and a reduction ratio of
$T_{\rm co}(\Omega=0.5 t)/T_{\rm co}(\Omega = 0) \approx 0.25$ has
been reported.\cite{freericks93,freericks00}

Especially in $d = \infty$ considerable effort has been expended to
construct weak-coupling perturbation theories which continuously
connect the adiabatic to the antiadiabatic
limit.\cite{freericks93,freericks00,freericks94a,freericks98} One
important outcome of this work was the suggestion that future research
should focus on generalizations of the ME theory that work with
dressed phonons (renormalized ME) but include higher order
non-adiabatic effects such as vertex corrections. Our method is such
an approach. A large number of studies based on the
Lang-Firsov\cite{lang63} method have been published (see
e.g.~Ref.~\onlinecite{alexandrov95}). For one dimensional systems the
Lang-Firsov results are in agreement with previous work.\cite{zheng89}
However, this is an uncontrolled approximation optimized for the
single polaron problem and its applicability to metallic densities is
not established.

In this paper we construct an essentially analytic theory of the
charge density wave transition temperature in the physically relevant
limit $\gamma \ll 1$. We build on previous work which showed how to
construct an expansion in $\gamma$ within the dynamical mean-field
approximation\cite{deppeler00} and which analysed the classical limit
($\Omega = 0$) of the charge density wave problem.\cite{blawid00} Here
we show how to calculate $T_{\rm co}$ including the first non-Migdal
term in $\gamma$ for arbitrary values of $\Omega/T_{\rm co}$. We find
that $T_{\rm co}$ depends very strongly on $\Omega$, becoming very
small as $\Omega$ is increased (at fixed coupling) beyond the
$\Omega=0$ value of $T_{\rm co}$, but that the specific behavior
depends strongly on whether spin-1/2 or spinless electrons are
considered. For spinless electrons, relevant to half-metals such as
CMR manganites, a divergent isotope exponent is obtained in the weak
electron-phonon coupling limit.

\section{Model and Formalism}

In this paper we study the simplest model of electrons interacting
with phonons, namely the Holstein model\cite{holstein59} 
$H_{\rm Hol} = H_{\rm el} + H_{\rm ph} + H_{\rm el-ph}$
with
\begin{eqnarray}
H_{\rm el} & = & -\sum_{ij} t_{i-j} \, 
c^{\dagger}_{i\sigma} c^{\phantom{\dagger}}_{j\sigma}
- \mu \sum_i 
\left(c^{\dagger}_{i\sigma} c^{\phantom{\dagger}}_{i\sigma} - n 
\right) \;, \\  
H_{\rm ph} & = & \frac{1}{2 \Lambda} \sum_i \left( r^{\, 2}_i 
+ \dot{r}_i^2/\Omega^2 \right) \;, \\
H_{\rm el-ph} & = & \sum_i r_i 
\left(c^{\dagger}_{i\sigma} c^{\phantom{\dagger}}_{i\sigma} - n \right)
\end{eqnarray}
where we have absorbed the electron-phonon coupling into the phonon
coordinate $r$ which thus has dimension of energy as does the phonon
stiffness parameter $\Lambda$. We further specialize to a mean
electron density per spin direction of $n = 1/2$; this implies $\mu =
0$. We also assume a bipartite lattice, which for our purposes we
define as a lattice possessing a dispersion $\epsilon_{\vec{k}}$
(Fourier transform of $t_{i-j}$) and a wavevector $\vec{Q}$ such that
$\epsilon_{\vec{k}-\vec{Q}} = -\epsilon_{\vec{k}}$ for all $\vec{k}$
in the Brillouin zone. An equivalent definition is that the lattice
may be divided into two sublattices $A$ and $B$, such that $t_{i-j}$
connects $A$ sites only to $B$ sites. The important feature of the
dispersion is the bare density of states per site per spin
$\rho(\epsilon )=\frac{1}{N}\,\sum_{\vec{k}}\delta (\epsilon-\epsilon_
{ \vec{k}})$. In this paper we use the semicircular form
$\rho(\epsilon) = \frac{1}{2 \pi t^{2}}\sqrt{4 t^{2}-\epsilon ^{2}}$.
It is useful to introduce the dimensionless electron-phonon coupling
$\lambda$ and adiabatic parameter $\gamma$ by
\begin{equation}
\lambda = \Lambda \rho_0,\;\;\; \gamma = \Omega \rho_0\;. 
\end{equation}
Here $\rho_0 = 1/\pi t$ denotes the bare density of states at the
chemical potential.

We now establish notation by reviewing the derivation of the $T_{\rm
co}$ equation within dynamical mean-field theory (DMFT). The
fundamental assumption of DMFT is the momentum independence of the
electron self energy $\Sigma_n = \Sigma(i\,\omega_n)$ ($\omega_n =
2\pi T\,(n+1/2)$). Consequently, in the absence of spatial symmetry
breaking, $\Sigma_n$ can be derived from an effective impurity model
specified by the action
\begin{eqnarray}
S(\{r_{nm}\};\{c_n\}) & = & (T/2\Lambda)\,
\sum_n r_{n0}\left( \omega_{n0}^2/\Omega^2+1 \right)\, r_{0n}
\nonumber\\
& & \!\!\!\!\!\!\!\!\!\!\!\!\!\!\!\!\!\!\!\!\!\!\!\!\!\!\!\!\!\!
-n\,\sum_n r_{n0} 
-(2s+1)\,{\rm Tr\,\,ln}\left[ c_n\,\delta_{nm} - T\,r_{nm} \right]\;.
\end{eqnarray}
In this action $r_{nm}$ are bosonic fields describing the phonons and
$c_n$ mean field functions describing the fermions (which have been
integrated out). We have indexed the bosonic fields by the difference
of two fermionic Matsubara frequencies $\omega_{nm} =
\omega_n-\omega_m$.  The factor $(2s+1)$ takes the values one and two
for spinless ($s=0$) and spin-1/2 ($s=1/2$) electrons,
respectively. The partition function is the functional integral over
the bosonic fields
\begin{equation}
Z = \int {\cal D}[r]\exp(-S)
\end{equation}
and is a functional of the effective field $c$ alone. It is useful to
define the impurity Green's function ${\cal G}$ and self energy
$\Sigma$ by 
\begin{equation}
\label{field}
{\cal G}_n \equiv \frac{1}{2s+1}\,\frac{\delta\,\ln Z}{\delta\,c_n} \equiv
\frac{1}{2s+1}\,\frac{1}{c_n-\Sigma_n(\{c_n\})}\;.
\end{equation}
The effective field is fixed by equating the local Green's functions of
the original lattice model and the effective impurity model. For a
semicircular density of states one obtains
\begin{equation}
c_n = i\omega _{n}+\mu -t^2\,{\cal G}_n \left( \{c_n\} \right)\;.
\end{equation}

For temperatures below the CDW transition the self energy acquires two
components for the two sublattices $A$ and $B$ which calls for two
effective impurity models given by the effective fields $a$ and $b$
respectively. The resulting self-consistency equations are:
\begin{eqnarray}
\label{dmft}
a_n &=& i\omega _{n}+\mu -t^2\,{\cal G}_n \left( \{b_n\} \right)\\
b_n &=& i\omega _{n}+\mu -t^2\,{\cal G}_n \left( \{a_n\} \right)\;.
\end{eqnarray}
Near the transition (assumed to be second order) we can expand $a_n =
c_n+\epsilon_n'$ and $b_n = c_n-\epsilon_n'$ leading to $\epsilon_n' =
t^2 \sum_m \partabl{{\cal G}_n}{c_m}\,\epsilon_m'$. Making use
of Eq.~(\ref{field}) we obtain
\begin{equation}
\label{tco2a}
\epsilon_n' = \frac{1}{1+(c_n-\Sigma_n)^2/t^2}\,
\sum_m \frac{{\rm \partial}\Sigma_n}{{\rm \partial} c_m}\,
\epsilon_m' \;.
\end{equation}
It is convenient to rewrite Eq.~(\ref{tco2a}) in the more symmetric
form\cite{blawid00,freericks94a}
\begin{equation}
\label{tco2}
\epsilon_n = -\sum_m \chi_n^{1/2}\,\Gamma_{nm}\,\chi_m^{1/2}
\,\epsilon_m \;.
\end{equation}   
Here we have introduced $\chi_n = -{\cal G}_n/(i\omega_n-\Sigma_n)$,
the irreducible two particle vertex $\Gamma_{nm} = \sum_{\sigma'}{\rm
\partial}{\Sigma_n^\sigma}/{\rm \partial}{\cal G}_m^{\sigma'}$ and
$\epsilon_n = \epsilon_n'\,\chi_n^{1/2}$.  Eq.~(\ref{tco2}) is the
statement that the matrix $ -\chi_n^{1/2} \Gamma_{nm} \chi_m^{1/2}$
has eigenvalue one and $T_{\rm co}$ is the temperature at which this
eigenvalue problem has a solution.  It can be easily
shown\cite{freericks94a} that $\chi_n$ is the `bare' CDW susceptibility
of the original lattice model $\chi_n = -1/N\,\sum_{\vec
k}\,G_n(\vec{k}+\vec{Q})\,G_n(\vec{k})$.
 
\section{Solution in the adiabatic limit}

\begin{figure}
\centerline{\psfig{file=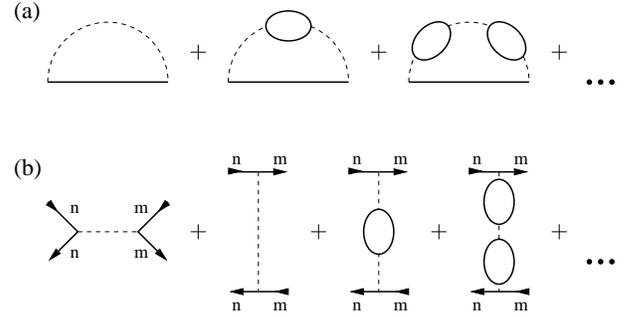,width=8cm,angle=0}}
\vspace*{0.1cm}
\caption{\label{dia.fig} \footnotesize Diagrammatic expansion of the
one-particle self-energy (a) and of the irreducible vertex of the
charge-order susceptibility (b) including all terms of order
$\bar{\gamma}$.  Heavy lines: full local Green's function ${\cal
G}_n$. Dashed lines: free phonon propagator. All other
possible diagrams are smaller by additional factors of
$\bar{\gamma}$ and are not shown.}
\end{figure}

As written, Eq.~(\ref{tco2}) is general, but not very useful because
not only must the matrix be diagonalized numerically, but also
determination of the matrix elements for arbitrary $\lambda$ and
$\gamma$ requires performing the functional integral over phonon
fields numerically (i.e.~via quantum Monte Carlo) and this becomes
prohibitively lengthly for physically relevant regimes of $\Omega$
and $T$. Here we turn this difficulty to advantage, exploiting the
smallness of $\Omega$ and $T$ to construct an analytic expansion in
powers of $\gamma$. The expansion builds on the
observation\cite{deppeler00} that the biggest contribution to a
diagram (e.g. for the electronic self energy or the vertex) with $n$
phonon loops is of order $[\max(T,\Omega)/t]^n$. This can be seen as
follows. A generic diagram $D^{(n)}$ with $n$ phonon loops contains
$n$ internal sums over bosonic Matsubara frequencies
$\omega_1,\ldots,\omega_n$ (here $\omega_n = \omega_{n0}$) each
associated with one (or more) free phonon propagators given by
\begin{equation}
{\cal D}^0_n = -\Lambda\,
\frac{\Omega^2}{\Omega^2+\omega_n^2} \;.
\end{equation}
If $F(i\omega_1,\ldots,i\omega_n)$ depicts the appropriate product of
Green's functions and phonon propagators we can write $D^{(n)} =
T^n\,\sum_{1,\ldots,n}\,F(i\omega_1,\ldots,i\omega_n)$. In the
classical limit, $\Omega/T \rightarrow 0$, the free phonon propagators
become delta functions and therefore $D^{(n)} \sim T^n$ as noted in
Ref. \onlinecite{blawid00}. In the quantum case it is more appropriate
to rescale the temperature $\tilde{T}=T/\gamma$ leading to $D^{(n)} =
\gamma^n \tilde{T}^n\,\sum_{1,\ldots,n}\,
F(i\gamma\tilde{\omega}_1,\ldots,i\gamma\tilde{\omega}_n)$. The
rescaled bosonic Matsubara frequencies $\tilde{\omega}$ become
continuous variables in the extreme quantum limit $\tilde{T}
\rightarrow 0$ and because $F$ has a small frequency expansion we
conclude that the biggest contributions to the $n$ phonon loop diagram
are of order $\gamma^n$ as argued in Ref. \onlinecite{deppeler00}. The
diagrammatic expansion shown in Fig.~\ref{dia.fig} includes all terms
with the minimal number of internal phonon loops, i.e. includes the
leading nontrivial order for the electronic self energy and the
vertex. Note that the leading order in $\gamma$ contains both, the
direct (first term in Fig.~\ref{dia.fig} (b)) and the exchange diagrams
for the vertex. The conventionally defined ME
approach\cite{marsiglio90,freericks93} only sums the bubble diagrams
for the CDW susceptibility, i.e.~the exchange diagrams for the vertex
(which we find to be of the same order) are neglected. Freericks and
co-workers\cite{freericks94a,freericks98,freericks94} have considered
a variety of different vertex corrections to the conventionally
defined ME approximation but they organized these in a coupling
constant expansion and not in the small frequency expansion we
considered here. Specific vertex corrections\cite{freericks94a} are of
higher order in $\gamma$ than the ones treated in this work.

Adding the electron bubbles in Fig.~\ref{dia.fig} leads to a `dressed'
phonon propagator ${\cal D}_{n} = [({\cal
D}_{n}^0)^{-1}-\Pi_{n}]^{-1}$ with a phonon self energy $\Pi_{n}$
given by
\begin{equation}
\Pi_{n} = (2s+1) T\,\sum_m {\cal G}_{n+m} {\cal G}_m \;.
\end{equation}
The occurrence of the free phonon propagator restricts the bosonic
Matsubara frequencies $\omega_{n}$ to values on the scale of
$\Omega$. Following the arguments given in
Ref. \onlinecite{deppeler00} we assume that the {\it dominant}
contributions to $\Pi_{n}$ come from frequencies of order $t$ and
consequently set the index $n$ under the sum to zero.\cite{deppeler00}
To leading order in $\gamma$ we can replace the full by the free Green's
function and find $\Pi_{n} = -1/\Lambda_c+{\cal O}(T^2)$ with
$\Lambda_c = 3/[4(2s+1)\rho_0]$.\cite{deppeler00,blawid00} The
`dressed' phonon propagator reads
\begin{equation}
{\cal D}_n = -\bar{\Lambda}\,
\frac{\bar{\Omega}^2}{\bar{\Omega}^2+\omega_n^2}\;
\end{equation}
with the renormalized phonon stiffness and phonon frequency
\begin {eqnarray}
\bar{\Lambda} & = & \Lambda/(1-\Lambda/\Lambda_c) \\
\bar{\Omega} & = & \Omega\,(1-\Lambda/\Lambda_c)^{1/2}\;.
\end{eqnarray}
Due to this influence of the electrons on the phonons the given
expansion is actually an expansion in $\bar{\lambda} = \bar{\Lambda}
\rho_0$ and $\bar{\gamma} = \bar{\Omega} \rho_0$. Maybe a more
familiar way to express the dressed electron phonon coupling is
$\bar{\lambda} = \rho_0\,\Lambda\,{\cal D}(0)$. The importance of not
neglecting the renormalization of phonons by electrons was stressed by
several authors, both in studies of $d=2$\cite{marsiglio90,noack} and
$d=\infty$\cite{freericks93,millis96} spatial dimensions. In the limit
$\Lambda \rightarrow \Lambda_c$ the renormalized phonon stiffness
$\bar{\Lambda} \rightarrow \infty$ and the renormalized phonon
frequency $\bar{\Omega} \rightarrow 0$ signaling that the lattice
becomes unstable against {\it local} distortions. In the absence of
long range order the ground state expectation value of the lattice
distortion $\langle r \rangle$ would change from $\langle r \rangle
=0$ for $\Lambda < \Lambda_c$ to $\langle r \rangle \neq 0$ for
$\Lambda > \Lambda_c$. This polaronic instability has been extensively
discussed in Ref. \onlinecite{millis96} and it is important to note
that the coupling constant is strongly enhanced and the adiabatic
parameter is strongly suppressed near the instability.

Within the small $\gamma$ limit we obtain
\begin{eqnarray}
\label{vertex}
\Gamma_{nm} & = & \frac{T}{\rho_0}\,\left[
-(2s+1)\lambda + \bar{\lambda}\,
\frac{1}{1+\tilde{\omega}_{nm}^2}
\right]\\
& & + {\cal O}\left( {\rm max}[T^2/t^2,\bar{\gamma}^2] \right)
\nonumber
\end{eqnarray}
and 
\begin{eqnarray}
\label{self}
\Sigma_n & = & \frac{\bar{\lambda}\,T}{\rho_0}\,
\sum_m {\cal G}_m\,
\frac{1}{1+\tilde{\omega}_{nm}^2} \\
& & + {\cal O}\left( {\rm max}[T^2/t^2,\bar{\gamma}^2] \right) \;.
\nonumber
\end{eqnarray}
Here, we have rescaled the bosonic frequencies $\tilde{\omega}_{nm} =
\omega_{nm}/\bar{\Omega}$. In Eq.~(\ref{self}) as well as in $\chi_n$,
we can replace the full Green's function by the bare one ${\cal G}_n
\rightarrow {\cal G}_n^0=[i\omega_n-i\,{\rm
sgn}(\omega_n)\sqrt{\omega_n^2+4 t^2}]/2t^2$ to leading order in
$\bar{\gamma}$. Previous studies\cite{ciuchi99} asserted that for
small phonon frequencies vertex corrections such as the second term in
Eq.~(\ref{vertex}) can be neglected. We see that this is not quite correct:
rather, the vertex corrections for $\bar{\gamma} \rightarrow 0$ affect
only the diagonal elements of the matrix in Eq.~(\ref{tco2}), but in
this diagonal region they are of the same order as the contribution from
the first part of Eq.~(\ref{vertex}).

The importance of the vertex corrections has been demonstrated in the
classical limit\cite{blawid00}. For $\bar{\gamma} \rightarrow 0$
($\bar{\Omega} \rightarrow 0$) we can expand
\begin{equation}
\label{class}
\frac{1}{1+\tilde{\omega}_{nm}^2}
\approx
\delta_{nm} +
\frac{1}{\tilde{\omega}_{nm}^2}\,
\left(1-\delta_{nm}\right) \rightarrow \delta_{nm} \; ,
\end{equation}
i.e.~the phonon propagator becomes diagonal for $\bar{\gamma} = 0$. In
this limit both vertex and self energy corrections to the BCS case
modify the upper cutoff of the logarithmic divergence in the CDW
susceptibility leading to a suppression of the CDW transition
temperature, which has been interpreted in terms of inelastic
scattering of electrons by phonons.\cite{blawid00}

Returning now to the quantum problem we insert Eq.~(\ref{vertex}) into the
eigenvalue equation (\ref{tco2}) yielding
\begin{eqnarray}
\label{tco}
\epsilon_n & = & \chi_n^{1/2}\,(2s+1)
\frac{\lambda\,T}{\rho_0}\,\sum_m \chi_m^{1/2}\,\epsilon_m\\
& &  -\chi_n^{1/2}\,\frac{\bar{\lambda}\,T}{\rho_0}\,
\sum_m \frac{1}{1+\tilde{\omega}_{nm}^2}
\chi_m^{1/2} \epsilon_m \nonumber \; .
\end{eqnarray}
For generic values of $\Omega/T_{\rm co}(0)$ we solve Eq.~(\ref{tco})
numerically by restricting the sums to a finite range $m < N$ and
diagonalizing the resulting matrix exactly numerically. Convergence is
improved by using an analytical approximation for Eq.~(\ref{tco}) for
large frequencies $\omega_n > \omega_N$. In the large frequency limit
the sums in Eq.~(\ref{self}) and Eq.~(\ref{tco}) including the phonon
propagator equal their $n=m$ values times the factor $T \sum_m
1/(1+\tilde{\omega}^2_m) = (\bar{\Omega}/2)\,\coth
(\bar{\Omega}/2T)$. Therefore, the coupling constant $(2s+1)\lambda$
can be replaced by the effective coupling constant
\begin{equation}
\label{effstrengtha}
\lambda^* = (2s+1)\lambda/[1-(2s+1)\lambda\,I)]
\end{equation} 
with
\begin{equation}
\label{effstrength}
I = \frac{T}{\rho_0}\,\sum_{|n|>N}\,
\frac{1}
{(\chi_n^0)^{-1}+
(\bar{\lambda}\,\bar{\Omega}/\rho_0)\,\coth
\left(\frac{\bar{\Omega}}{2\,T}\right)}
\end{equation}
where we have introduced $\chi_n^0 = -{\cal G}_n^0/i\omega_n$. For
$n,m = 0,\ldots,N$ the leading eigenvalue of the matrix
\begin{eqnarray}
\label{matrix}
M_{nm} & = &
\frac{2\lambda^*T}{\rho_0}\,\chi_n^{1/2}\,\chi_m^{1/2}
-\frac{\bar{\lambda}T}{\rho_0}\,\chi_n^{1/2}\,\chi_m^{1/2}\\
& & \!\!\!\!\!\!\!\!\!\!\!\!\!\!\!\!\!\!\!\!\!\!
\times \left\{
\frac{1}{1+[2\pi (n-m)T/\bar{\Omega}]^2} +
\frac{1}{1+[2\pi (n+m)T/\bar{\Omega}]^2}
\right\}\nonumber
\end{eqnarray}
has to be found. $T_{\rm co}$ is given by the condition that the
leading eigenvalue equals one. We have checked the convergence with $N$;
typically we used values ranging from $N = 256$ to $N = 4096$ yielding
a relative error $\delta T_{\rm co}/T_{\rm co} < 0.01$. For later use
we give the weak coupling $\bar{\lambda} \rightarrow 0$ result for the
effective coupling constant. In this limit the charge density wave
transition temperature becomes very small and the Matsubara sum can be
evaluated as an integral yielding $ 1/\lambda^* \approx
1/[(2s+1)\lambda] - \ln(2 t/\omega_N)$.

\section{Results}

\begin{figure}
\centerline{\psfig{file=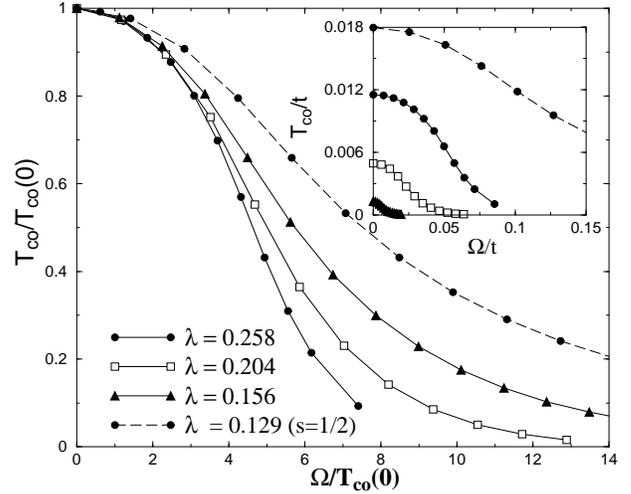,width=8cm,angle=270}}
\caption{\label{tco.fig} \footnotesize CDW transition temperature as
function of the typical phonon frequency $\Omega \sim M^{-1/2}$ for
spinless electrons. In the main panel $T_{\rm co}$ and $\Omega$ are
scaled to the zero frequency value of the CDW transition temperature
$T_{\rm co}(0)$. Circles connected with dashed lines represent results
for $\lambda = 0.129$ but spin $s=1/2$. Inset: $T_{\rm co}$ as
function of $\Omega$ on the scale $t$.}
\end{figure}
The CDW transition temperature resulting from a numerical
diagonalization of (\ref{matrix}) is shown in Fig.~\ref{tco.fig}.
Decreasing the ionic mass, i.e.~increasing the phonon frequency
$\Omega$, is seen to suppress $T_{\rm co}$ considerably for both
spinless and spin-1/2 electrons. The suppression becomes very
pronounced when $\Omega \sim (2s+1) \pi T_{\rm co}(\Omega=0)$. This is
in qualitative agreement with the strong suppression of the zero
temperature order parameter reported in $d=1$ where the CDW state is
destroyed when the phonon frequency is of the order of the adiabatic
($\gamma = 0$) mean-field gap value.\cite{bursill98} We note that the
scaled $T_{\rm co}$ curves fan out at larger phonon frequencies: in
the extreme quantum limit the suppression of $T_{\rm
co}(\Omega)$ relative to $T_{\rm co}(\Omega=0)$ depends strongly on
coupling. This is an indication that quantum non-Migdal effects (high
energy electron self energy $\Sigma_n \approx 1/2\, \bar{\Lambda}
\bar{\Omega}\,{\cal G}_n^0$) are different from classical non-Migdal
effects ($\Sigma_n \approx \bar{\Lambda} T \,{\cal G}_n^0$).

$T_{\rm co}(0)$ increases considerably for spin $s=1/2$, essentially
because the spin sum increases the bare susceptibility by a factor of
two. Nevertheless, the change in spin $s$ cannot be accounted for by a
change of $\lambda$ (i.e. $\rho_0$) only. Reducing $\lambda$ by a
factor of two when changing the spin from $s=0$ to $s=1/2$ does not
bring the transition temperature curves close to each other as shown
in the inset of Fig.~\ref{tco.fig} (see the circles connected with a
solid, $s=0$, or dashed, $s=1/2$, line). The rather different
suppression of $T_{\rm co}$ for $s=0$ and $s=1/2$ was also found in
studies in $d=1$.\cite{hirsch83,bourbonnais89} Even on the scale
$T_{\rm co}(0)$ the large frequency tails are qualitatively different in
the two cases. We provide some analytic insight in Section~V.

Within the spin-1/2 Holstein model sizable transition temperatures
might be obtained even in the antiadiabatic limit and for modest
coupling strengths making the CDW transition accessible to QMC
simulations. Freericks and Jarell report a suppression of $T_{\rm
co}/T_{\rm co}(0) \approx 0.25$ for $\Omega/T_{\rm co}(0) \approx 7.1$
(corresponding to $\gamma = 0.282$).\cite{freericks93} This result,
obtained numerically in the antiadiabatic limit, indicates an even
stronger suppression of $T_{\rm co}$ than obtained here in the
adiabatic limit ($\gamma$ values used in our calculation vary from 0
to 0.08) but is of the same order. The comparison of absolute numbers
is difficult because the work Ref. \onlinecite{freericks93} does not
use the semicircular density of states we have chosen.

\begin{figure}
\centerline{\psfig{file=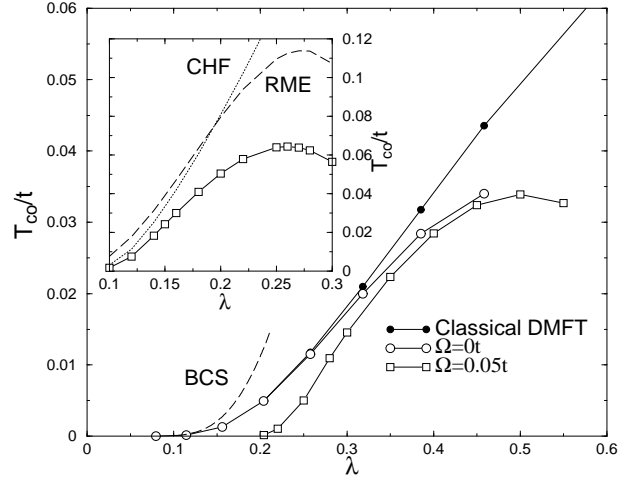,width=8cm,angle=270}}
\caption{\label{opt.fig} \footnotesize Charge density wave transition
temperature as function of the coupling strength $\lambda$ for $s=0$,
calculated from Eq.~(\protect{\ref{tco}}) for $\gamma = 0$ (white
circles, identical to the modified BCS approximation of Ref.~9) and
$\gamma = 0.05/\pi$ (white squares).  Also shown are the `BCS'
approximation\protect{\cite{blawid00}} (dashed line) and a numerical
solution of the classical DMFT equations.\protect{\cite{blawid00}} For
$\lambda=0.35$ ($\bar{\lambda} \bar{\gamma} = 7.6 \times 10^{-3}$)
differences (amplified because $T_{\rm co}$ depends exponentially on
parameters) between the exact classical treatment and our approximate
results become noticeable, suggesting that the expansion breaks down.
The inset presents data for $s=1/2$ and $\gamma = 0.1/\pi$. The
solution of Eq.~(\protect{\ref{tco}}) (white squares) is compared to
the renormalized Migdal-Eliashberg theory (RME) and to the conserving
Hartree-Fock approximation (CHF) discussed in Ref.~14. The
$\bar{\gamma}$ expansion can be trusted up to $\lambda = 0.18$ which
corresponds to $\bar{\lambda} \bar{\gamma} = 7.9 \times 10^{-3}$. For
details see text.}
\end{figure}
It is important to note that $T_{\rm co}(0)$ is not the mean-field CDW
transition temperature $T_{\rm co}^{\rm MF} =
(8t\gamma/e\pi)\,\exp[-1/(2s+1)\lambda]$.\cite{blawid00} Even at zero
phonon frequency the charge density wave transition temperature is
suppressed by thermal fluctuations of the phonon fields, e.g. $T_{\rm
co}(0)/T_{\rm co}^{\rm MF} = 0.33, 0.40, 0.49$ for $s=0$ and
$\lambda=0.258, 0.204, 0.156$ respectively.  In Fig.~\ref{opt.fig}
$T_{\rm co}$ is plotted as function of the coupling constant
$\lambda$. The suppression of the charge density wave transition
temperature relative to its `BCS' or static mean field value due to
thermal fluctuations (white circle) and due to additional quantum
fluctuations (white squares) is clearly visible. At weak coupling
quantum fluctuations of the phonon fields are so effective in reducing
$T_{\rm co}$ that it seems unlikely that CDW materials can be found
with a Debye temperature $\Theta_{\rm D}$ much larger than the mean
field transition temperature $T_{\rm co}^{\rm MF}$. Higher order
corrections in $\bar{\gamma}$ are accompanied by higher powers of
$\bar{\lambda}$ and can no longer be neglected when the coupling
constant is large. The comparison with the full numerical solution of
the classical DMFT equation given in Ref. \onlinecite{blawid00}
reveals that the expansion suggested here breaks down already at the
surprisingly small value of $\bar{\gamma} \bar{\lambda} \approx
10^{-2}$. The inset of Fig.~\ref{opt.fig} will be discussed in Section
VI.

\begin{figure}
\centerline{\psfig{file=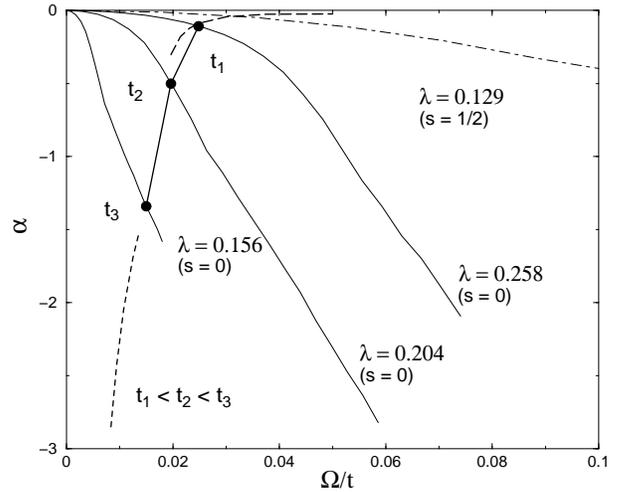,width=8cm,angle=270}}
\caption{\label{iso.fig} \footnotesize Isotope exponent as function of
$\Omega/t$ for spinless electrons. With increasing phonon frequency
$\alpha$ drops to huge negative values. The circles connected with a
thick solid line represents the change in $\alpha$ with increasing $t$
for fixed values of $\Lambda = 0.49$ and $\Omega =
0.015$. ($t_1=1.65$, $t_2=1.31$, $t_1=1$) The dashed and dotted line
are extrapolations given by Eq.~(\protect{\ref{isolo}}) and
Eq.~(\protect{\ref{isohi}}), respectively. In the latter we have used a
cutoff parameter $\eta = 1.16$. The isotope exponent for spin-1/2
electrons (dot-dashed line) remains finite.}
\end{figure}
The dependence of the CDW transition temperature on the ionic mass is
usually discussed in terms of the isotope exponent $\alpha =
\frac{1}{2}\,\frac{\Omega}{T_{\rm co}}\, \frac{{\rm d}T_{\rm co}}{{\rm
d}\Omega}$ which is shown in Fig.~\ref{iso.fig}. Because of the rapidly
increasing factor $\Omega/T_{\rm co}$ the negative isotope exponent
$-\alpha$ monotonically increases to very large values for spinless
electrons. In the case $s=1/2$, however, the isotope exponent flattens
off as function of $\Omega/t$ due to the different large frequency
behavior of $T_{\rm co}$. 

\section{Analytic expressions}

In the adiabatic and extreme quantum limits we can derive explicit
analytic expressions for $T_{\rm co}$. First, we consider the limit
$\Omega/T_{\rm co}(\Omega=0) \rightarrow 0$, i.e.~we expand
(\ref{tco}) around the classical solution obtained in
Ref. \onlinecite{blawid00}. An appropriate Ansatz for the eigenvector
$\epsilon_n$ is
\begin{equation}
\label{eigen}
\epsilon_n = {\cal C}\,
\frac{\chi_n^{1/2}}{\left[1+(\bar{\lambda}\,T/\rho_0)\,\chi_n\right]}\;.
\end{equation}
Note that, $\epsilon_n$ is not quite the classical solution because it
still depends on $\bar{\gamma}$ via $\chi_n$. Therefore, after
inserting (\ref{eigen}) into (\ref{tco}) we still have to expand in
$\bar{\gamma}$ which finally leads to
\begin{equation}
\label{lotco}
\frac{1}{2s+1} = \frac{\lambda\,T}{\rho_o}\,
\sum_n \frac{\chi_n^{\rm class}}{1+(\bar{\lambda}\,T/\rho_0)\,\chi_n^{\rm
class}}
-\lambda\,K\,\left(
\frac{\Omega}{T_{\rm co}(0)}
\right)^2
\end{equation}
with $\chi_n^0 = (\sqrt{(2 t/\omega_n)^2+1}-1)/2 t^2$. $K$ is a
numerical coefficient independent of temperature but weakly dependent
on $\bar{\lambda}$. Its magnitude is relatively small, e.g.  $K =
1.775\times10^{-2}$ for $\lambda = 0.156$ ($s=0$). 

The first part of Eq.~(\ref{lotco}) is the classical $T_{\rm co}$
equation.\cite{blawid00} The sum has a logarithmic $T$
dependence. Therefore, expanding $T = T_{\rm co}(0) - \delta T$ leads
to a term $[\lambda/T_{\rm co}(0)]\,\delta T$. Consequently, the CDW
transition temperature in the limit $\Omega/T_{\rm co}(0) \ll 1$ is
given by
\begin{equation}
\label{lotcosol}
\frac{T_{\rm co}}{T_{\rm co}(0)} = 1 - 
K\,\left(\frac{\Omega}{T_{\rm co}(0)}\right)^2\;.
\end{equation}
The quadratic drop is clearly seen in Fig.~\ref{tco.fig}. We can use
Eq.~(\ref{lotcosol}) to derive the isotope exponent in the classical
limit. We obtain
\begin{equation}
\label{isolo}
\alpha = -K\,\left(\frac{\Omega}{T_{\rm co}(0)}\right)^2
= \frac{T_{\rm co}}{T_{\rm co}(0)} -1\;.
\end{equation}
The isotope exponent as function of $t$ according to this result is
depicted in Fig.~\ref{iso.fig} as a dashed line. The exponential decrease
of $T_{\rm co}(0)$ with increasing kinetic energy $t$ of the electrons
is responsible for the sharp upturn of the negative isotope exponent.  

In the extreme quantum limit the temperature is not a relevant high
energy scale. Therefore, the upper cutoff frequency introduced above
should be given by $\omega_N = \eta\,\bar{\Omega}$ to yield results to
logarithmic precision with $\eta$ being a numerical factor of order
one. We can replace the phonon propagator by
\begin{equation}
\label{quant}
\frac{1}{1+\tilde{\omega}_{nm}^2}
\approx
1-\tilde{\omega}_{nm}^2 \;.
\end{equation}
Moreover, to the same precision all Matsubara sums can be replaced by
integrals and the temperature occurs only as a lower cutoff in the
theory. The $T_{\rm co}$ equation (\ref{tco}) reads simply
\begin{equation}
\label{hitco}
1 = \left(\lambda^*-\bar{\lambda}\right)\,
t\,\int_{T}^{\eta\,\bar{\Omega}}\,\chi(\omega) \,{\rm d}\omega\;.
\end{equation}
For small frequencies the self energy (\ref{self}) is given by
$\Sigma_n = - i\omega_n\,\bar{\lambda}$. Therefore, $\chi(\omega) =
\chi_0(\omega)/(1+\bar{\lambda})$ which, inserted in
Eq.~(\ref{hitco}), gives the CDW transition temperature in the limit
$\Omega/T_{\rm co} \gg 1$
\begin{equation}
\label{hitcosol}
T_{\rm co} = \eta\,\bar{\Omega}\, \exp\left( 
-\frac{1+\bar{\lambda}}{\lambda^*-\bar{\lambda}} \right) 
\equiv \eta\,\bar{\Omega}\,\exp\left(-\frac{1}{\lambda_{\rm eff}}\right)\;.
\end{equation}
Note that, $\bar{\Omega}$ enters the calculation as the upper cutoff
of the integral (\ref{hitco}) and as lower cutoff in the effective
interaction strength $\lambda^*$ (see (\ref{effstrengtha})) in close
resemblance to the well known McMillan formula for the superconducting
transition temperature.\cite{mcmillan68} Because the effective
coupling constant $\lambda_{\rm eff}$ depends logarithmically on
$\bar{\Omega}$ (via $\lambda^*$) we obtain for the isotope exponent in
the extreme quantum limit
\begin{equation}
\label{isohi}
\alpha = \frac{1}{2}\,
\left[1-\frac{1+\bar{\lambda}}{(1-\bar{\lambda}/\lambda^*)^2}\right]\;.
\end{equation}
The extreme quantum limit for fixed values of the phonon frequency
$\Omega$ is realized for small values of $\lambda$ (or large values of
$t$) due to the fast drop of $T_{\rm co}(0)$ with decreasing coupling
strength (see Fig.~\ref{opt.fig}). For spinless electrons
the isotope exponent becomes $\alpha \approx -1/(2\,\lambda^2\,I^2)$,
i.e.~$\alpha$ diverges in the limit $\lambda \rightarrow 0$ with
power law behavior. The divergence of the negative isotope exponent
with increasing kinetic energy of the electrons is shown in
Fig.~\ref{iso.fig} as a dotted line. The isotope exponent for spin-1/2
electrons, however, remains finite, equalling $\alpha = -3/2$ for
$\lambda \rightarrow 0$. The difference between the cases $s=0$ and
$s=1/2$ is due to a near-cancellation of the vertex parts for
$s=0$. Consequently, the leading order of $\lambda_{\rm eff}$ is
${\cal O}(\lambda^2)$ for $s=0$ and ${\cal O}(\lambda)$ for $s=1/2$.

In the ME theory for the superconducting transition temperature the
effect of Coulomb interaction is considered by introducing a
pseudopotential $\mu^*$.\cite{mcmillan68} Therefore, it is tempting to
absorb Coulomb effects on the charge density wave transition
temperature in a similar way in $\lambda^*$. Indeed, a repulsive
Coulomb interaction is an electron-electron interaction similar to the
attractive vertex part, first term in Fig.~\ref{dia.fig}(b), and will
change the value of $\lambda^*$ simply by replacing $(2s+1) \lambda
\rightarrow (2s+1) \lambda-U$ (see also
Ref. \onlinecite{freericks95}). $\lambda^*$ acts as a pseudopotential
with a non vanishing value for $U=0$. Nevertheless, Coulomb
interaction will also change the electron self energy and the `vertex
corrections' (second and following terms in Fig.~\ref{dia.fig}(b))
which is a more subtle effect due to the different frequency
dependence of electron-phonon and Coulomb interactions and is left for
future work.

\section{Relation to traditional approaches}

It is traditional to write the CDW transition temperature in the form
$T_{\rm co} = W_{\rm eff}\,\exp\left(-1/\lambda_{\rm eff}\right)$
defining an effective bandwidth and coupling constant. The effective
parameters $W_{\rm eff}$ and $\lambda_{\rm eff}$ are usually obtained
by fitting to experimental data and therefore the functional
dependence on the microscopic properties like electron-phonon coupling
and adiabatic parameter remains hidden. The isotope exponent, however,
explores this dependence and tests our physical understanding of the
CDW transition.

A `BCS' or static mean field approach\cite{gruner94} would result in
$T_{\rm co}^{\rm MF} = 2t \exp[-1/(2s+1)\lambda]$, i.e. $T_{\rm co}$
depends exponentially on $\lambda$ but is independent of $\gamma$. For
strong coupling the `BCS' theory is clearly wrong because the Holstein
model in this limit can be mapped on an effective pseudospin
model\cite{hirsch83} and $T_{\rm co}$ has to decrease with increasing
$\lambda$. Here we have demonstrated that also the weak coupling
behavior is not predicted correctly and that instead of an `BCS'
isotope exponent $\alpha = 0$ an enhanced ($s=1/2$) or even divergent
($s=0$) negative isotope effect is obtained.

Corrections to the BCS approximation have been considered. In the
conventional Migdal-Eliashberg (ME) theory\cite{migdal58} of
electron-phonon effects in solids electron self energies arising from
the interaction with the phonons are included but all vertex
corrections are neglected. Other authors\cite{marsiglio90,freericks93}
have argued that one should include also the feedback of the electrons
on the phonon system but still neglecting the vertex corrections, thus
defining a renormalized Migdal-Eliashberg theory (RME). In our
notation this corresponds to dropping the second ($\bar{\lambda}$)
term in Eq.~(\ref{tco}). The RME approximation predicts an incorrect
isotope effect because $T_{\rm co}$ depends on ion mass only via the
dependence of $\chi_m$ on the normal state self energy, and comparison
to our results shows that this dependence is not the leading one. At
very weak coupling the RME theory reduces to the BCS theory (up to
logarithmic corrections) and predicts $\alpha = 0$. Freericks and
co-workers added extra vertex corrections to the RME
theory\cite{freericks94a} but these are higher order in $\gamma$ then
the effects we studied in this paper.  

We have plotted the charge density wave transition temperature for
various coupling strengths $\lambda$ and fixed adiabatic parameter
$\gamma = 0.1/\pi$ and $s=1/2$ in the inset of Fig.~\ref{opt.fig} for
both the RME approach and the present work. The charge density wave
transition temperature obtained within RME theory is always larger
than the one obtained from the systematic $\bar{\gamma}$ expansion
presented here.  A similar plot for a larger value of $\gamma =
0.5/\sqrt{\pi}$ has been presented in Ref. \onlinecite{freericks93},
where the RME results were compared with QMC data, showing that the
RME theory indeed overestimates the transition temperature for weak
coupling strength at least in the antiadiabatic limit. Note that, the
RME results presented in Ref. \onlinecite{freericks93} are not
precisely identical to ours because we have included the phonon self
energy only to leading order in $\bar{\gamma}$, but we believe that
this does not effect the conclusions drawn here. It is well known that
$T_{\rm co}(\lambda)$ is maximized at a $\lambda_{\rm max} <
\infty$. [$T_{\rm co} \rightarrow 0$ as $\lambda \rightarrow 0$
because the coupling vanishes; $T_{\rm co} \rightarrow 0$ as $\lambda
\rightarrow \infty$ because in this limit a picture of almost
completely localized particles with intersite interactions $\sim
t^2/\lambda$ applies.] The present approach underestimates both
$\lambda_{\rm max}$ and $T_{\rm max}$, as may be seen
(Fig.~\ref{opt.fig}) by comparison to the classical treatment of
Ref. \onlinecite{blawid00} (to which the present treatment reduces as
$T_{\rm co}$ becomes greater than $\Omega$) and by comparison to QMC
data (Ref. \onlinecite{freericks93}) and indeed is not expected to
apply for $\lambda \sim \lambda_{\rm max}$.

A theory neglecting the renormalization of the phonon propagator by
electrons predicts that $T_{\rm co}$ is maximized at $\lambda = \infty$.
As an example we have plotted in the inset of Fig.~\ref{opt.fig} the
conserving Hartree-Fock approximation discussed by Freericks {\it et
al.}\cite{freericks93} This corresponds to using Eq.~(\ref{tco}) but
with $\bar{\lambda}$ replaced by $\lambda$. In the weak coupling limit
the difference between $\bar{\lambda}$ and $\lambda$ is negligible, so
the CHF theory gives a good estimate of the CDW transition
temperatures. The comparison with QMC data given in
Ref. \onlinecite{freericks93} demonstrate that the CHF theory, and
therefore the theory given here, can (at least for $s=1/2$ and weak
coupling strengths) be extended to the antiadiabatic limit.

Finally, we point out that the systematic expansion in $\bar{\gamma}$
discussed in this work reproduces the results of the renormalized ME
theory when applied to the calculation of the superconducting
transition temperature $T_c$. Following Freericks and
Scalapino\cite{freericks94} we find $T_c \approx \eta\,\bar{\Omega}
\exp[-(1+\bar{\lambda})/\bar{\lambda}]$ in the extreme quantum
limit. For the half-filled Holstein model the charge density wave
transition temperature is always larger than the superconducting
transition temperature if $\Omega < \infty$; only as $\Omega
\rightarrow \infty$ do they become equal.\cite{ciuchi99} The
superconducting transition temperature is controlled by a logarithmic
divergence with upper cutoff set by the phonon frequency whereas the
corresponding upper cutoff in the theory for $T_{\rm co}$ is enhanced
towards $2t$ by the first diagram in Fig.~\ref{dia.fig}(b).

\section{Conclusion and Comparison to Experiment}

We have used the dynamical mean field method to obtain a systematic
and tractable treatment of the isotope effect on charge density wave
transitions. Our method is applicable to the physically relevant case
in which both phonon frequency and transition temperature are small
relative to the electronic bandwidth, and in combination with the
results of Ref. \onlinecite{deppeler00} may be generalized to the crucial
case in which other interactions beyond the electron-phonon one are
important.

Quantum phonon effects are found to suppress the charge density wave
transition temperature severely. This is in agreement with previous
work on one dimensional models demonstrating that the suppression of
$T_{\rm co}$ is not due to the strong quantum fluctuations
characteristic of one spatial dimension. The isotope exponent of
$T_{\rm co}$ for spin-1/2 electrons is enhanced compared to the ME
prediction. In the case of spinless electrons the isotope exponent may
even be divergent. Although beyond the scope of the theory, the
consideration of phonon self energy effects lead to an optimal charge
density wave transition temperature for values of the coupling
strength slightly below the polaronic instability at $\lambda_c$ but
peak position and height are underestimated.

Our results suggest that it will be difficult to find CDW systems with
mean field transition temperatures much less than typical Debye
frequencies. It is of interest to examine data with this in mind.  For
quasi-one-dimensional CDW materials the {\it mean field} CDW
transition temperature is typically larger than room temperature and
therefore larger than or of the same order as the Debye temperature
$\Theta_{\rm D}$. In $\rm TaS_3$, $\rm K_{0.3}MoO_3$ and $\rm
(TaSe_4)_2I$ one deduces experimentally values of $T_{\rm co}^{\rm MF}
\approx 540K$, $T_{\rm co}^{\rm MF} \approx 330$ and $T_{\rm co}^{\rm
MF} \approx 1040K$ respectively.\cite{gruner94} Quasi-two-dimensional
CDW materials include the dichalcogenides and the molybdenum bronzes
and oxides. For these materials it is more difficult to deduce mean
field values of the CDW transition temperature because the CDW
transition is not accompanied by a metal insulator transition. For
$2H-{\rm TaSe_2}$ one might argue from measurements of the optical
conductivity\cite{vescoli98} for a value of $T_{\rm co}^{\rm MF}
\approx 580K$ similar to the quasi-one-dimensional materials.  Charge
order transitions are also observed in many two and three dimensional
transition metal oxides including some members of the high-temperature
superconductor and `CMR' family.\cite{imada98} Frequently, the Coulomb
repulsion of the charge carriers is responsible for the charge order
transition and therefore beyond the scope of this paper.
Nevertheless, the phenomenology is very similar and e.g.~for $\rm
La_{1.67}Sr_{0.33}NiO_4$ one might deduce $T_{\rm co}^{\rm MF} =
750K$. As was suggested by our theory the case of $\Theta_{\rm D} \gg
T_{\rm co}^{\rm MF}$ seems to be extremely rare.

Our results bear at least a qualitative similarity to data obtained on
`CMR' manganites. Recently, a colossal isotope shift of the CDW
transition temperature in $\rm Nd_{0.5}Sr_{\rm 0.5}MnO_3$ was
reported.\cite{zhao98} With magnetic field the negative isotope
exponent varies from $\sim 1.2$ to $\sim 4.5$. Within the double
exchange picture of the manganites an increase of the applied magnetic
field means an increase of the kinetic energy $t$. Due to the large
Hunds rule coupling of the conduction electron spins to the core spins
of the Mn ions the conduction electrons are fully spin
polarized. Therefore we should compare to the case $s=0$. To
demonstrate the effect of an increasing $t$ in our calculation we draw
a line for fixed $\Lambda$ and $\Omega$ in Fig.~\ref{iso.fig}. Indeed,
we find a monotonically increasing negative isotope effect with
increasing $t$.\\

{\it Acknowledgements.} We thank A.~Deppeler and M.~Gugros for useful
discussions. SB acknowledges the DFG, the Rutgers University Center
for Materials Theory and NSF DMR0081075 for financial support; AJM
acknowledges NSF DMR0081075.


\end{document}